\documentclass[12pt]{article}
\usepackage{esfconf}
\usepackage{epsfig}
\begin{document}

\title{The phasetransition in the multiflavour Schwinger model}

\authors{S.\,D\"urr\adref{1}}

\addresses{\1ad Paul Scherrer Institut, Particle Theory Group,
5232 Villigen PSI, Switzerland}

\maketitle

\begin{abstract}
A summary is given of a quantization of the multiflavour Schwinger model on
a finite-temperature cylinder with chirality-breaking boundary conditions at
its spatial ends, and it is shown that the analytic expression for the chiral
condensate implies that the theory exhibits a second order phasetransition with
$T_c\!=\!0$.
\end{abstract}



\bigskip
\noindent
The (euclidean) Schwinger model (SM), i.e.\ QED(2) with one or several massless
fermions, is defined through the generating functional
\begin{equation}
Z[\bar\eta,\eta]\sim\int\!D(A,\bar\psi,\psi)\;
e^{-\int{1\over4}F^2-\int\bar\psi D\!\!\!\!/\;\psi+
\int(\bar\eta\psi+\bar\psi\eta)}
\label{SMgenfunc}
\end{equation}
where the integration is over all smooth fields on the manifold $M$, e.g.\ a
torus $T_{L,\beta}\!=\![0,L]\times[0,\beta]$. The theory allows for a
finite-temperature interpretation if bosonic (fermionic) fields are
(anti-)periodic in the second direction and $\beta\ll L$.
From (\ref{SMgenfunc}) one may proceed to compute an observable
\begin{equation}
\langle O\rangle_{L,\beta}={1\over Z[0,0]}
\int\!D(\mathrm{fields})\;O(\mathrm{fields})\;e^{-S(\mathrm{fields})}
\label{SMobservable}
\end{equation}
and study its behaviour at various box-lengths $L$ and temperatures
$T\!=\!1/\beta$.

From a mathematical point of view the choice $M\!=\!T_{L,\beta}$ looks
appealing, since the finite volume provides a natural infrared regularization,
gauge-fields fall into different topological classes labeled by 
$\nu\!=\!{g\over4\pi}\int\!F_{\mu\nu}\epsilon_{\mu\nu}$, and the index-theorem
holds true \cite{Joos}: The number of exact zero-modes of the Dirac operator on
a given background $A$ equals $|\nu|$, and the nonzero modes come in pairs
($D\!\!\!\!/\,\psi_\pm\!=\!\pm\mathrm{i}\lambda\psi_\pm$) of opposite chirality
($\gamma_5\psi_\pm\!=\!\pm\psi_\pm$), i.e.\ the diagonal entries (in a chiral
representation) of the Green's function on the subspace orthogonal to the
zero-modes vanish: $S'_{\pm\pm}\!=\!\mathrm{tr}(P_\pm S')\!=\!0$, where
$P_\pm\!=\!{1\over2}(1\!\pm\!\gamma_5)$. This and the form (\ref{SMgenfunc})
takes after the fermions have been integrated
out~\cite{SachsWipf}
\begin{equation}
Z[\bar\eta,\eta]\sim\sum_{\nu\in\mathrm{\bf Z}}\int\!DA^{(\nu)}\;\,
e^{-\int{1\over4}F^2}\,\,
\prod_{k=1}^{N_{\!f}|\nu|}(\bar\eta\psi_k)(\bar\psi_k\eta)\;
{\det}'(D\!\!\!\!/\,)^{N_{\!f}}\;e^{\int\bar\eta S'\eta}
\label{SMtorus}
\end{equation}
(the product is over the zero-modes !) imply for the chiral condensate on
the torus
\begin{equation}
\langle\bar\psi P_\pm\psi\rangle_{L,\beta}=
{\delta^2\over\delta\bar\eta_\pm\delta\eta_\pm}
\Big\vert_{\bar\eta,\eta\!=\!0}\log Z[\bar\eta,\eta]\;
\left\{{\neq0\;(N_{\!f}\!=\!1)\atop=0\;(N_{\!f}\!\geq\!2)}\right.
\;.
\label{SMcondensate}
\end{equation}
In the singleflavour case the well-known (anomaly induced) Schwinger value
$g e^\gamma/4\pi^{3/2}$ is reproduced for $L,\beta\to\infty$.
The problem is that in the multiflavour theory the condensate on the torus is
zero already at finite $L,\beta$ and not only in the limit
$L\beta\to\infty$ (where it has to vanish, due to Coleman's theorem
\cite{Coleman}); this makes the torus, from a physical point of view, an
uninteresting manifold.

\vspace*{-1mm}
\begin{figure}[b]
\begin{tabular}{lr}
\epsfig{file=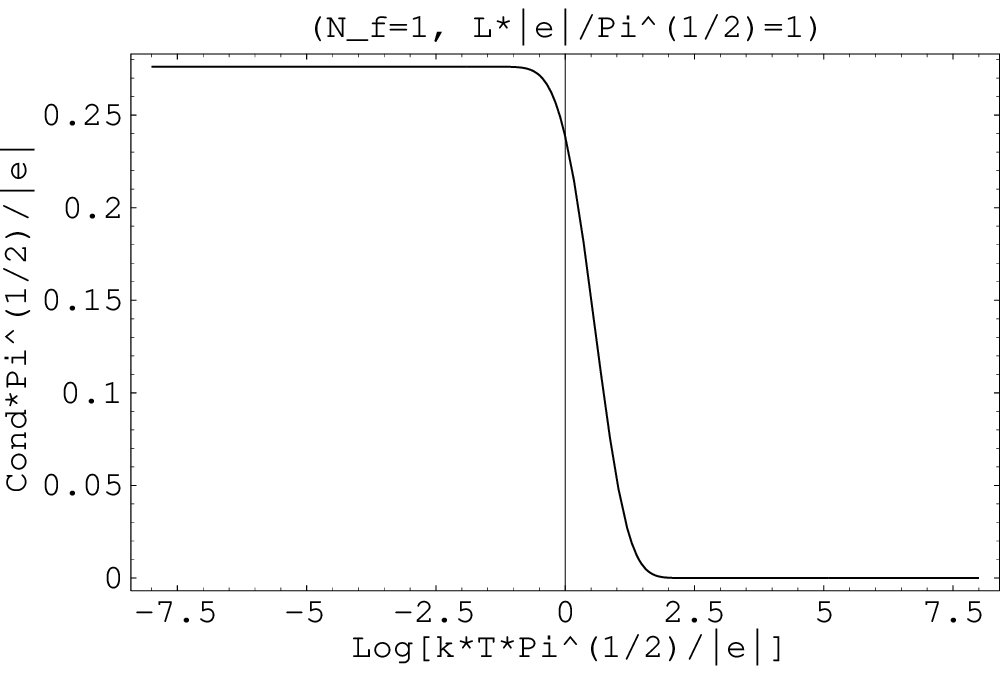,width=6.5cm}&
\epsfig{file=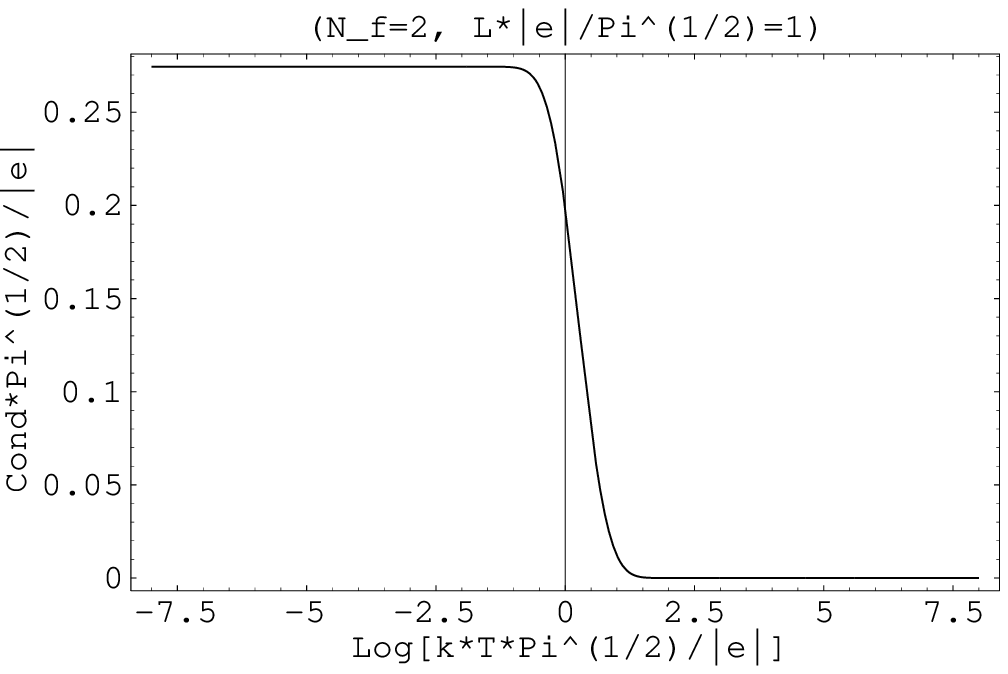,width=6.5cm}\\
\end{tabular}
\vspace*{-2mm}
\caption{Crossover-phenomenon of the dimensionless condensate
$\vert\langle\bar\psi P_\pm\psi\rangle\vert/\mu$
as a function of $\log(kT/\mu)$ at fixed $L\!=\!1/\mu$
for $N_{\!f}\!=\!1$ (left) and $N_{\!f}\!=\!2$ (right).}
\label{figtwo}
\end{figure}

Quite generally in field theory and statistical mechanics, if one wants to
establish that a system breaks a global symmetry spontaneously (or investigate
how it avoids doing so), one {\em needs to break the symmetry explicitly\/} and
observe how the system behaves as the symmetry breaking source is {\em turned
off\/}. A straightforward way of doing this, both in QCD and the SM, is to give
the fermions a mass $m$. The problem with this choice, in the case of the SM,
is that it destroys exact analytic tractability\footnote{Bosonization rules
\cite{BosonizationRules} allow to separate the theory into a heavy (above the
mass-gap) and a light (below) sector, and recently an exact solution for the
latter has appeared \cite{SineGordonExact}.}. Alternatively, one may break the
chiral symmetry by boundary conditions and study implications of sending the
boundaries to infinity \cite{HraskoBalogWipfDurrDurrWipf}.
The idea is to quantize the SM on a thermal cylinder $[0,L]\times[0,\beta]$,
where at the two ends $x^1\!=\!0,L$ (no identification) fields are subject to a
member of a one-parameter family of local linear boundary conditions (labeled
by $\theta\!\in\!\mathrm{\bf R}$).
Peculiar properties of this choice include \cite{HraskoBalogWipfDurrDurrWipf}:
\newline${}\;\bullet\,$
No $U(1)_V$ current leaks through the boundary, i.e.\ $j_\perp\!=\!\psi^\dagger
n\!\!\!/\,\psi\!=\!0$ on $\partial M$.
\newline${}\;\bullet\,$
The topological charge is {\em not quantized\/}, i.e.\ $\nu={g\over4\pi}\int
\epsilon_{\mu\nu}F_{\mu\nu}={g\over2\pi}\int E\in\mathrm{\bf R}$.
\newline${}\;\bullet\,$
There is no pairing and Dirac modes have no fixed chirality,
i.e.\ $(S_\theta)_{\pm\pm}\!\neq\!0$.
\newline${}\;\bullet\,$
The boundary condition ($\theta$) adds a P/CP-odd piece to the effective
action%
\footnote{Note that in 2 dimensions the couping $g$ has the dimension of a
mass.}
\begin{equation}
\Gamma=
{1\over2g^2}\int\phi(\bigtriangleup^2-N_{\!f}{g^2\over\pi}\bigtriangleup)\phi+
\theta{N_{\!f}\over2\pi}\int\bigtriangleup\phi+
N_{\!f}\Gamma(c)\;,
\label{SMeffaction}
\end{equation}
where the Hodge decomposition
$gA_\mu\!=\!-\epsilon_{\mu\nu}\partial_\nu\phi\!+\!
\partial_\mu\chi\!+\!{2\pi\over\beta}c\,\delta_{\mu2}$
has been used.
\newline${}\;\bullet\,$
There are no fermion zero-modes, hence (\ref{SMtorus}) is replaced by the
simpler form
\begin{equation}
Z[\bar\eta,\eta]\sim\int\!D\phi\,dc\;
e^{-\Gamma}
e^{\bar\eta S_\theta\eta}\;.
\label{SMcylinder}
\end{equation}
For $N_{\!f}\!=\!1$ and $N_{\!f}\!=\!2$ the resulting functional integral
may be evaluated completely \cite{Durr}. Specializing to midpoints
($\xi\!=\!{L\over2}$), the condensate for $N_{\!f}\!=\!1$ is
\begin{equation}
\begin{array}{rcl}
\!\!\!\!{\langle\bar\psi P_{\pm}\psi\rangle\over\mu}\!&\!=\!&\!
\pm{e^{\pm\theta/\mathrm{cosh}(\lambda/2)}\over4\pi}
\Big(1+2\sum\limits_{n\geq1}(-1)^n\;
{1\over\mathrm{cosh}(n\pi\tau)}\:\exp(-n^2\pi\tau)\Big)\,\cdot
\\
&{}&\!
\exp\Big\{
\gamma-2\sum\limits_{j\geq1}(-1)^jK_0(j\lambda)
\Big\}\cdot
\hfill(\tau\!\gg\!1)
\\
&{}&\!
\exp\Big\{\sum\limits_{n\geq0}
{4\over(2n\!+\!1)(e^{2(2n\!+\!1)\pi\tau}-1)}-
((2n\!+\!1)\!\rightarrow\!\sqrt{(2n\!+\!1)^2\!+\!({\lambda\over\pi})^2\,})
\Big\}
\\
\!\!\!\!{\langle\bar\psi P_{\pm}\psi\rangle\over\mu}\!&\!=\!&\!
\pm{e^{\pm\theta/\mathrm{cosh}(\lambda/2)}\over4\pi}
\sum\limits_{m\geq0}(\!-1)^m\;
{e^{-\pi((2m+1)^2-1)/4\tau}\over\mathrm{sinh}(\pi(2m+1)/2\tau)}\times
\\
&{}&\!
\sum\limits_{k\geq0}\mathrm{cosh}(\mbox{${\pi(2m+1)(2k+1)\over2\tau}$})
({\rm erf}(\mbox{${(k+1)\sqrt{\pi}\over\sqrt{\tau}}$})
\!-\!{\rm erf}(\mbox{${k\sqrt{\pi}\over\sqrt{\tau}}$}))\,\cdot
\\
&{}&\!
\exp\Big\{\gamma+{\pi(1-\mathrm{tanh}(\lambda/2))\over\sigma}
-2\sum\limits_{j\geq1}K_0(j\sigma)
\Big\}\cdot
\hfill(\tau\!\ll\!1)
\\
&{}&\!
\exp\Big\{
-\sum\limits_{m\geq1}
{2\over m(e^{m\pi/\tau}+1)}
-(m\!\rightarrow\!\sqrt{m^2\!\!+\!\!({\sigma\over2\pi})^2\,})
\Big\}
\end{array}
\label{SMcondone}
\end{equation}
while for $N_{\!f}\!=\!2$ the result reads
\begin{equation}
\begin{array}{rcl}
\!\!{\langle\bar\psi P_{\pm}\psi\rangle\over\mu}\!&\!=\!&\!
\pm{2^{1/4}e^{\pm\theta/\mathrm{cosh}(\lambda/\sqrt{2})}\over
4\sqrt{\pi\;}\sqrt{\lambda\;}}\,\cdot
\\
&{}&\!
\Big(1+2\sum\limits_{n\geq1}(-1)^n\;
{e^{-n^2\pi\tau/2}\over\mathrm{cosh}(n\pi\tau)}\cdot
{\sum\limits_{k\in\mathrm{\bf Z}}
{e^{-(n/2-k)^22\pi\tau}+e^{-(n/2+k)^22\pi\tau}\over2}\over
\sum\limits_{k\in\mathrm{\bf Z}}e^{-2k^2\pi\tau}}\Big)\,\cdot
\\
&{}&\!
\exp\Big\{
{\gamma\over2}-\sum\limits_{j\geq1}(-1)^j K_0(j\sqrt{2}\;\lambda)
\Big\}\cdot
\hfill(\tau\!\gg\!1)
\\
&{}&\!
\exp\Big\{\sum\limits_{n\geq0}
{2\over(2n\!+\!1)(e^{2(2n\!+\!1)\pi\tau}-1)}-
((2n\!+\!1)\!\rightarrow\!\sqrt{(2n\!+\!1)^2\!+\!2({\lambda\over\pi})^2\,})
\Big\}
\\
\!\!{\langle\bar\psi P_{\pm}\psi\rangle\over\mu}\!&\!=\!&\!
\pm{2^{1/4}e^{\pm\theta/\mathrm{cosh}(\lambda/\sqrt{2})}\over
4\sqrt{\pi\;}\sqrt{\sigma\;}}\;
\sum\limits_{m\geq0}(-1)^m\;
{e^{-\pi((2m+1)^2-1)/8\tau}\over\mathrm{sinh}(\pi(2m+1)/2\tau)}\times
\\
&{}&\!
{
\sum\limits_{q\in Z}\!e^{-\pi q^2/2\tau}\!\sum\limits_{p\geq0}\!
e^{(-1)^{p+q+m}{\pi(p+1/2)(m+1/2)\over\tau}}
({\rm erf}({(p+3/2)\over\sqrt{2\tau/\pi\;}})
\!-\!{\rm erf}({(p-1/2)\over\sqrt{2\tau/\pi\;}}))
\over
\sum\limits_{q\in Z}e^{-\pi q^2/2\tau}
}\,\cdot
\\
&{}&\!
\exp\Big\{{\gamma\over2}
+{\pi(1\!-\!\mathrm{tanh}(\lambda/\sqrt{2}))\over2\sqrt{2}\sigma}
-\sum\limits_{j\geq1}K_0(j\sqrt{2\;}\sigma)
\Big\}\cdot
\hfill(\tau\!\ll\!1)
\\
&{}&\!
\exp\Big\{
-\sum\limits_{m\geq1}
{1\over m(e^{\pi m/\tau}+1)}\!-\!
(m\!\rightarrow\!\sqrt{m^2\!+\!2({\sigma\over2\pi})^2\,})
\Big\}
\end{array}
\label{SMcondtwo}
\end{equation}
where $\mu\!=\!{g\over\sqrt{\pi}}$ and $\tau\!=\!{\beta\over2L},
\sigma\!=\!\mu\beta, \lambda\!=\!\mu L$.
In either case a representation which converges fast for $\tau\!\gg\!1$ and
another one which converges fast for $\tau\!\ll\!1$ is given.
Note however, that for finite $L,\beta$ the two representations agree exactly.

\vspace*{-1mm}
\begin{figure}[t]
\epsfig{file=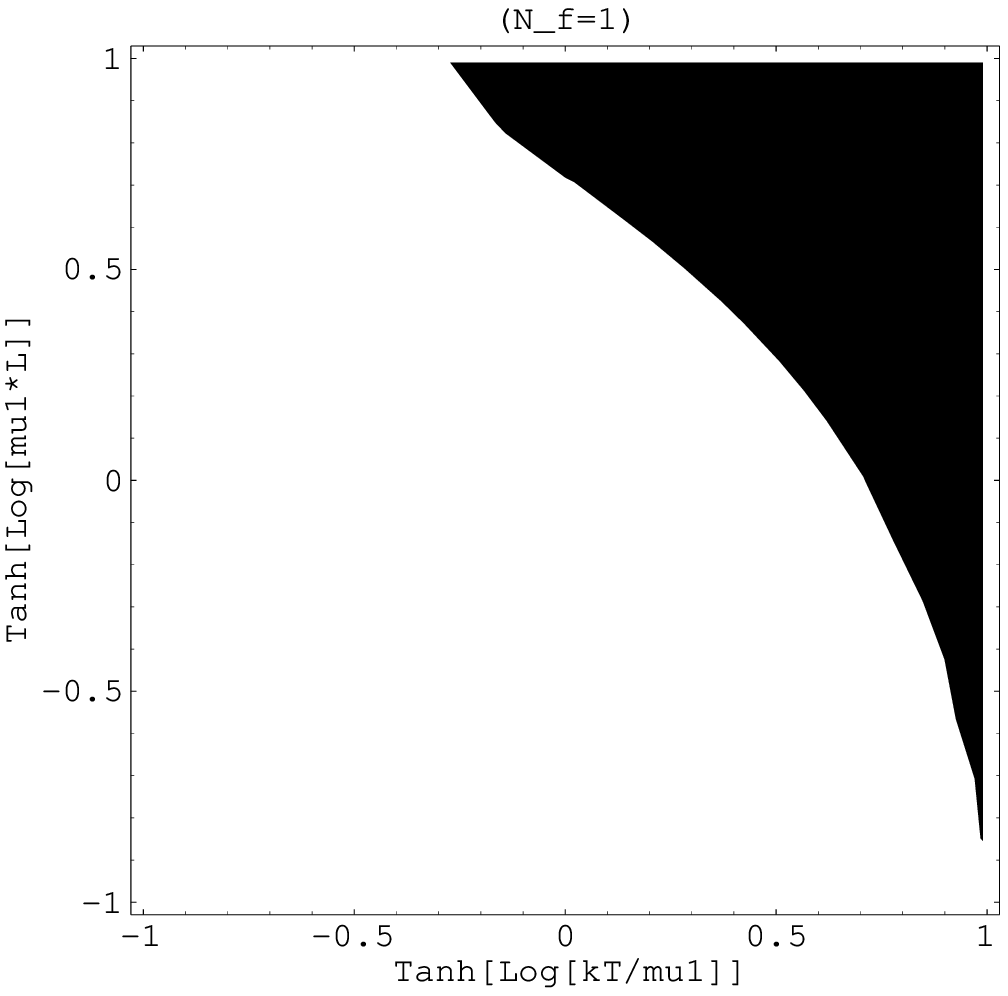,width=5.4cm}
\hspace*{2.6cm}
\epsfig{file=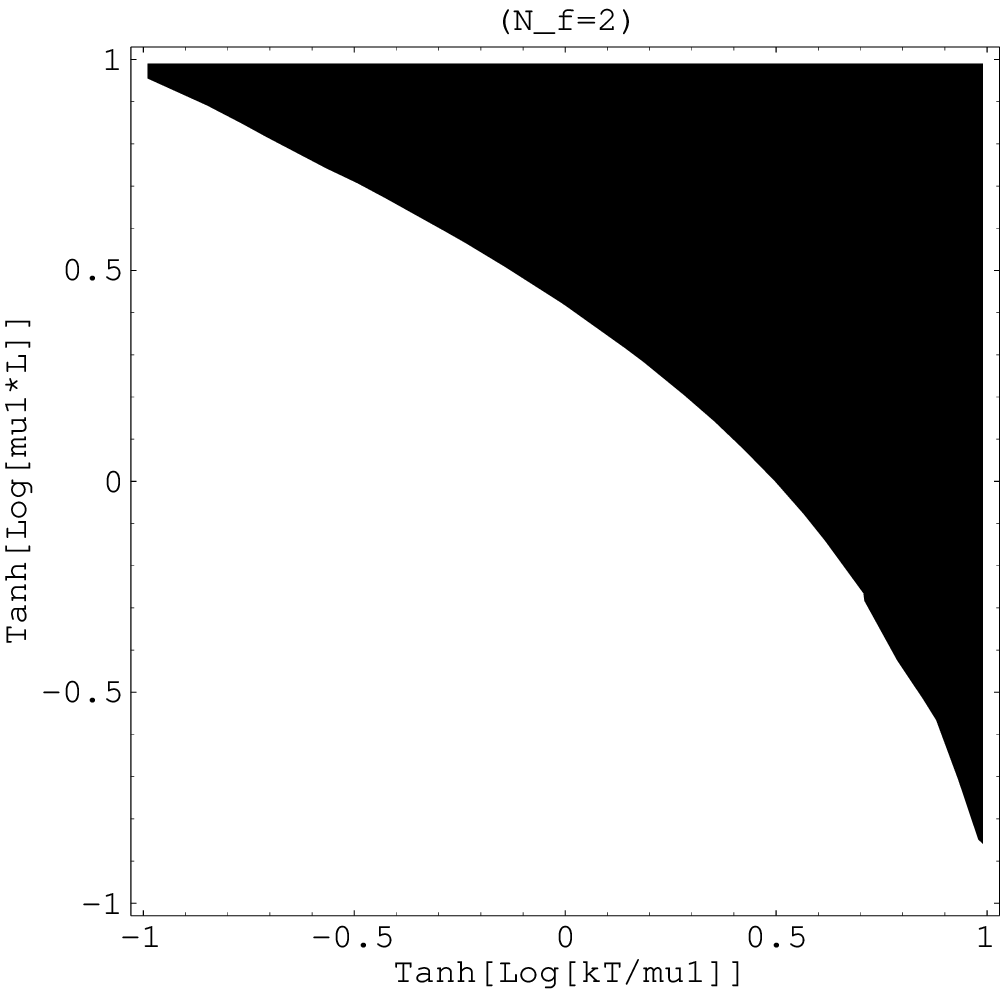,width=5.4cm}
\vspace*{-2mm}
\caption{Quasi-Phasestructure in the $\log(kT/\mu)$ - $\log(L\mu)$ plane
for $N_{\!f}\!=\!1$ and $N_{\!f}\!=\!2$.
White points: $\langle\bar\psi\psi\rangle/\mu\!>\!e^\gamma/8\pi$.
Black points: $\langle\bar\psi\psi\rangle/\mu\!<\!e^\gamma/8\pi$.}
\end{figure}

A numerical evaluation of (\ref{SMcondone}) and (\ref{SMcondtwo}) --~originally
undertaken just to check the latter assertion~-- gives a picture as shown in
Fig.~1:
For any finite~$L$ there is a low-temperature regime where the combination of
anomalous~($N_{\!f}\!=\!1$) or attempted spontaneous ($N_{\!f}\!=\!2$) and
explicit symmetry breaking generates a non-zero condensate, and there is a
``critical temperature'' where the condensate decays through a fairly well
localized ``symmetry quasi-restoration'' process to a value exponentially close
(bot not equal) to zero.
Repeating the exercise for various $L$ values \cite{Durr}, one realizes that
crossover-temperature and height of the plateau increase with diminishing $L$
both for $N_{\!f}\!=\!1$ and 2, whereas for $L\mu\!\to\!\infty$ remarkable
differences show up: For $N_{\!f}\!=\!1$ kink-position and plateau-height
are fairly stable, while for $N_{\!f}\!=\!2$ the kink keeps moving left and the
plateau-height decreases unlimitedly.
It makes thus sense to introduce the concept of a {\em quasi-phasestructure\/}
\cite{Durr} in order to distinguish points in parameter-space where the
condensate is {\em manifestly non-zero\/} (white in Fig.~2) from those where
the chiral symmetry is {\em quasi-restored\/} (black in Fig.~2).
The point $\beta\!=\!L\!=\!\infty$ (upper left corner in both plots) is clearly
in the broken (quasi-)phase for $N_{\!f}\!=\!1$, while it seems to lie near or
right on the crossover-line for $N_{\!f}\!=\!2$ (note that areas close to the
boundaries in Fig.~2 are cut off for numerical reasons).
A closer look at (\ref{SMcondone}, \ref{SMcondtwo}) reveals that indeed the
one-flavour condesate approaches its (finite) value at $\beta\!=\!L\!=\!\infty$
smoothly from any direction (within the plot), whereas in the two-flavour
theory the condensate {\em vanishes differently, depending on which limit is
performed first\/} \cite{Durr}:
\begin{equation}
\langle\bar\psi\psi\rangle
\sim
\left\{
\begin{array}{ll}
1/\sqrt{\lambda}&(N_{\!f}\!=\!2, T\!=\!0)\\
e^{-\mathrm{const}\,\lambda}&(N_{\!f}\!=\!2, T\!>\!0)
\end{array}
\right.
\;.
\label{SMsmellPT}
\end{equation}
Of course this ``smells'' like a critical phenomenon, but in order to make
contact with the general theory, we need to convert our result to an expression
where the symmetry is broken in the bulk, i.e.\ by a fermion mass term.
The naive (dimensionally motivated) identification $L^{-1}\!\leftrightarrow\!m$
then leads to ($N_{\!f}\!\geq\!2$)
\begin{equation}
\langle\bar\psi\psi\rangle
\sim
\left\{
\begin{array}{lc}
m^{(N_{\!f}-1)/N_{\!f}}\!&\!(T\!=\!0)\\
e^{-\mathrm{const}/m}\!&\!(T\!>\!0)
\end{array}
\right.
\Longrightarrow\;
\chi={d\over dm}\langle\bar\psi\psi\rangle
\sim
\left\{
\begin{array}{lcc}
m^{-1/N_{\!f}}\!&\!\rightarrow\!&\!\infty\\
{e^{-\mathrm{const}/m}\over m^2}\!&\!\rightarrow\!&\!0
\end{array}
\right.
\!\!\!,
\label{SMnaivePT}
\end{equation}
where we have generalized to arbitrary $N_{\!f}\!\geq\!2$ .
If (\ref{SMnaivePT}) were correct, it would indicate a second order
phasetransition with critical temperature $T_c\!=\!0$.
The problem is that the corresponding critical coefficient $\delta$ in
$\langle\bar\psi\psi\rangle\sim m^{1/\delta}$ (at the transition) would take
the value $N_{\!f}/(N_{\!f}\!-\!1)$, which is at variance with the correct
(bosonization rule based) result $\delta\!=\!(N_{\!f}\!+\!1)/(N_{\!f}\!-\!1)$ 
\cite{SmilgaVerbaarschot}.

The problem can be solved by realizing that the entire manifold
$[0,L]\times[0,\beta]$ our calculation was based on, may be seen as a model
of the inside of a two-dimensional meson at finite temperature (the chirality
breaking boundary conditions at $x^1\!=\!0,L$ do indeed resemble the MIT
bag boundary conditions), hence the proper identification is
$L^{-1}\!\leftrightarrow\!M$, where $M\!\sim\!m^{\#}$ is the mass of the
lightest (pseudoscalar) meson.
Unlike in QCD, the exponent $\#$ depends on the number of active flavours. 
The simplest way to get the value $\#$ is the following:
{\em Require\/} that the multiflavour SM satisfies, in complete analogy with
QCD, a Gell-Mann -- Oakes -- Renner type PCAC-relation%
\footnote{Admittedly, this is done with an eye on the bosonization approach,
where (\ref{SMGOR}) was first {\em derived\/} from the proper scaling of
$\langle\bar\psi\psi\rangle$ and $M$ \cite{Smilga}.}
(for $T\!=\!0, N_{\!f}\!\geq\!2$)
\begin{equation}
m\langle\bar\psi\psi\rangle\sim M^2,
\label{SMGOR}
\end{equation}
from which, upon using $\langle\bar\psi\psi\rangle\sim
(1/L)^{(N_{\!f}\!-\!1)/N_{\!f}}\sim(m^{\#})^{(N_{\!f}\!-\!1)/N_{\!f}}$,
one concludes
\begin{equation}
1+\#\,{N_{\!f}\!-\!1\over N_{\!f}}=2\#
\;\;\;\Longrightarrow\;\;\;
\#={N_{\!f}\over N_{\!f}\!+\!1}
\;.
\end{equation}
Hence, using the relationship
$L^{-1}\!\leftrightarrow\!m^{N_{\!f}/(N_{\!f}\!+\!1)}$, the corrected version
of (\ref{SMnaivePT}) is
\begin{equation}
\langle\bar\psi\psi\rangle
\sim
\left\{
\begin{array}{lc}
m^{(N_{\!f}-1)/(N_{\!f}+1)}\!&\!(T\!=\!0)\\
e^{-\mathrm{const}/m^{N_{\!f}/(N_{\!f}+1)}}\!&\!(T\!>\!0)
\end{array}
\right.
\;\Longrightarrow\;\;
\chi
\sim
\left\{
\begin{array}{lcc}
m^{-2/(N_{\!f}+1)}\!&\!\rightarrow\!&\!\infty\\
{e^{-\mathrm{const}/m^{N_{\!f}/(N_{\!f}+1)}}\over
m^{1+N_{\!f}/(N_{\!f}+1)}}\!&\!\rightarrow\!&\!0
\end{array}
\right.
\!\!,
\label{SMeducatedPT}
\end{equation}
from which we see that the multiflavour SM shows indeed a second order
phasetransition with zero critical temperature and the critical exponent
$\delta$, defined through $\langle\bar\psi\psi\rangle(T\!=\!T_c)\sim
m^{1/\delta}$, is
\begin{equation}
\delta={N_{\!f}\!+\!1\over N_{\!f}\!-\!1}
\;,
\end{equation}
as originally derived by Smilga and Verbaarschot \cite{SmilgaVerbaarschot}.

We shall conclude with two comments:

({\sl i\/})
The SM feature worth emphasizing is the analogy with QCD: For $N_{\!f}\!=\!1$
either theory shows a smooth crossover and the nonzero chiral condensate does
not indicate SSB, since the symmetry is already broken by the anomaly.
For $N_{\!f}\!\geq\!2$ and with a small symmetry breaking source there is a
striking simi\-larity to QCD {\em slightly above the phasetransition\/}:
The Polyakov loop in the multiflavour SM is real and positive, and the chiral
condensate is almost zero; the system ``tries'' to break the axial flavour
symmetry spontaneously, the spectrum shows a ``mass gap'' between the
``Schwinger particle'' with mass $g\sqrt{N_{\!f}\over\pi}+O(m)$
(which is the analogue of the $\eta'$) and the $N_{\!f}^2\!-\!1$ light
``Quasi-Goldstones'' with mass $M\sim m^{N_{\!f}/(N_{\!f}+1)}$
\cite{BosonizationRules, Smilga}.
The latter get sterile in the chiral limit (as required by Coleman's theorem
\cite{Coleman}), but the important point is that, as long as one stays away
from the chiral limit, these ``pions'' dominate the long-range Green's
functions between external (S,P,V,A)-currents.

({\sl ii\/})
The sketch presented above exemplifies the glory and the misery of a
path-integral quantization of a soluble model with non-standard boundary
conditions:
Symmetry breaking boundary conditions prove useful to force a field theory
into a definite groundstate and help, for this reason, to explore systems
which successfully show SSB or attempt doing so.
On the other hand, the path-integral approach is not very transparent; formulas
may be clumsy (cf.\ (\ref{SMcondone}, \ref{SMcondtwo})) and one has no clue
what are the relevant physical degrees of freedom.
It is therefore little surprise that the fact that the multiflavour SM shows
a second order phasetransition with zero critical temperature (the only $T_c$
possible in 2 dimensions) was first derived in the bosonized approach
\cite{BosonizationRules, SmilgaVerbaarschot}.


\end{document}